\documentclass[aps,preprint,nofootinbib,superscriptaddress]{revtex4}
\usepackage{amssymb}

\usepackage{epsfig}

\begin{document}

\date{\today}
\title{ Quasi-elastic scattering and fusion with a modified Woods-Saxon potential}

\author{Ning Wang}
\email{wangning@gxnu.edu.cn}  \affiliation{Department of Physics,
Guangxi Normal University, Guilin 541004, People's Republic of
China}

\author{Werner Scheid}
\affiliation{Institut f\"{u}r Theoretische Physik der
Universit\"{a}t, D-35392 Giessen, Germany}

\begin{abstract}

The elastic and large-angle quasi-elastic scattering reactions were
studied with the same nucleus-nucleus potential proposed for
describing fusion reactions. The elastic scattering angle
distributions of some reactions are reasonably well reproduced by
the proposed Woods-Saxon potential with fixed parameters at energies
much higher than the Coulomb barrier. With an empirical barrier
distribution based on the modified Woods-Saxon potential and taking
into account the influence of nucleons transfer, the calculated
quasi-elastic scattering cross sections of a series of reactions are
in good agreement with the experimental data.

\end{abstract}

\maketitle


\begin{center}
\textbf{I. INTRODUCTION}
\end{center}

Heavy-ion quasi-elastic scattering and fusion reactions at energies
around the Coulomb barrier have been extensively studied in recent
decades, since they provide an ideal opportunity to obtain the
information of nuclear structure and nucleus-nucleus interaction and
to explore the mechanism of heavy-ion reactions at near barrier
energies which is of great importance for the synthesis of
super-heavy nuclei
\cite{Muh08a,Hag04,Row91,Lei95,Tim95,Gup05,Zag01,Ada98,Liu06}. Based
on the quantum tunneling concept, it is thought that the
quasi-elastic scattering (a sum of elastic, inelastic scattering,
and transfer channels) is a good counterpart of the fusion reaction
in the sense that the former is related to the reflection
probability of a potential barrier while the latter is related to
the penetration probability \cite{Hag04}. In addition, it has been
shown that the fusion barrier distribution generated by the coupling
of the relative motion of the nuclei to internal degrees of freedom
can be extracted from precisely measured fusion excitation functions
\cite{Row91,Lei95}. The similarity of the barrier distribution can
be extracted from large-angle quasi-elastic scattering excitation
functions \cite{Tim95} which can be more easily measured than the
fusion excitation functions \cite{Tim97}. Therefore, it is expected
that both the fusion and quasi-elastic scattering cross sections of
a heavy-ion reaction at energies around the Coulomb barrier can be
unifiedly described by the same nucleus-nucleus potential. However,
in recently published papers \cite{Muk07,Muh08}, Mukherjee et al.
found that the Woods-Saxon nuclear potential can not simultaneously
reproduce precise fusion and elastic scattering measurements of
$^{12}$C+$^{208}$Pb \cite{Muk07}, and Zamrun and Hagino found that
the depth parameter of the Woods-Saxon potential for describing the
fusion cross sections of $^{16}$O+$^{144}$Sm have to be readjusted
to reproduce the experimental quasi-elastic scattering cross
sections of this reaction with the same coupled-channels framework
\cite{Muh08}. To solve this discrepancy, it is necessary to find a
nucleus-nucleus potential for a unified description of the
scattering and fusion data in heavy-ion reactions. In addition, for
giving satisfying predictions of quasi-elastic scattering cross
sections for unmeasured reaction systems, it is required to find a
nucleus-nucleus potential to describe quasi-elastic scattering
reactions systematically.

Studies of quasi-elastic scattering reactions and transfer
processes, especially of the behavior of the transfer probabilities
as functions of the distance of closest approach or the incident
energies have attracted a lot of attention. Some investigations show
that the semiclassical method is suitable for describing the
heavy-ion scattering at large reaction
distance\cite{Bass80,Rehm91,Liang93,Shr00}. The transfer probability
is expressed as an exponential function of the distance between the
reaction partners $P_{\rm tr} \propto  \exp(-2 \alpha R_c)$
\cite{Bass80} in the semiclassical approximation, where $\alpha$ is
the transfer form factor and $R_c$ is the distance of closest
approach between two nuclei. The exponential dependence on $R_c$ is
a characteristic property of tunneling \cite{Liang93}. At energies
below the barrier the experimental slopes are generally in good
agreement with the predictions of the model for one-nucleon
transfer. At higher energies, the measured slopes deviate from the
calculated values, which is often referred to as "slope anomaly" in
addition to other types of anomalies which are in connection with
slopes obtained in two particle transfer reactions \cite{Roy95}.
Some experiments show an energy dependence of the slopes, and a
clear trend of a decrease of slope parameters as a function of
increasing energy was found in \cite{Shr00,Roy95}. The transfer
probabilities at below barrier energies have been extensively
studied while a theoretical model for describing the slope
parameters and the transfer probabilities at energies near and above
the Coulomb barrier has not been well established yet. The study of
the transfer probability in the latter energy region is still
required. In addition, it is interesting to explore the relation
between the transfer probabilities and the barrier distribution
since the transfer probabilities generally peak in the vicinity of
the barrier energies \cite{Tim97}.

In \cite{Wang08} we proposed a modified Woods-Saxon potential model
based on the Skyrme energy-density functional together with the
extended Thomas-Fermi approach. This model was first proposed in
\cite{Liu06} and a large number of fusion reactions have been
described satisfactorily well with an empirical barrier distribution
which is based on the calculated entrance channel potential. In this
work, we try to describe the heavy-ion elastic and quasi-elastic
scattering with the same potential for describing the fusion
reactions. The paper is organized as follows: In Sec.II, the
theoretical model for the description of the elastic and
quasi-elastic scattering is introduced. In addition, some calculated
results are compared with experimental data. The summary and
discussion are given in Sec.III.

\newpage

\begin{center}
\textbf{II. THEORETICAL MODEL FOR ELASTIC, QUASI-ELASTIC SCATTERING
AND FUSION}
\end{center}

In this section, we first briefly introduce the modified Woods-Saxon
potential and the elastic scattering is studied with the potential.
In addition, the empirical barrier distribution is briefly
introduced for describing fusion reactions. Then, the quasi-elastic
scattering and the transfer probabilities are described with the
empirical barrier distribution. Some calculated results are also
presented in this section.

\begin{center}
\textbf{A.  Modified Woods-Saxon Potential and Elastic Scattering at
above Barrier Energies}
\end{center}

In \cite{Wang08} we proposed a Woods-Saxon potential model based on
the Skyrme energy-density functional together with the extended
Thomas-Fermi approach. The nucleus-nucleus interaction potential
reads as:
\begin{eqnarray}
V(R) =  V_N(R)+V_{C}(R).
\end{eqnarray}
Here, $V_N$ and $V_C$ are the nuclear and Coulomb interactions,
respectively. We take $V_C(R)=e^2 Z_1 Z_2/R$, and the nuclear
interaction $V_N$:
\begin{eqnarray}
V_N(R)=\frac{V_0}{1+\exp [(R-R_0)/a]},
\end{eqnarray}
with \cite{Dob03}
\begin{eqnarray}
V_0=u_0 [1+\kappa (I_1+I_2)]\frac{A^{1/3}_1 A^{1/3}_2}{A^{1/3}_1+
A^{1/3}_2},
\end{eqnarray}
and
\begin{eqnarray}
R_0=r_0(A^{1/3}_1+A^{1/3}_2)+c.
\end{eqnarray}
$I_1=(N_1-Z_1)/A_1$ and $I_2=(N_2-Z_2)/A_2$ in Eq.(3) are the
isospin asymmetries of the projectile and target nuclei,
respectively. In this potential, the depth of the potential $V_0$
depends on the reaction system and the isospin asymmetries. To
distinguish it from the traditional Woods-Saxon potential (with
three parameters) in which the depth of the potential is independent
of the reaction system and the isospin asymmetries, we call the
proposed potential (with five parameters) "modified" Woods-Saxon
potential. The parameters of the modified Woods-Saxon (MWS)
potential \cite{Wang08} are determined by the entrance channel
potentials of 66996 reactions obtained with the Skyrme
energy-density approach and are listed in Table 1.

\begin{table}
\caption{ Parameters of the modified Woods-Saxon potential.}
\begin{tabular}{cccccc}
\hline\hline
 & $ r_0 \, (\rm fm) $ & $c \, (\rm fm)$ & $u_0 \, (\rm MeV)$ & $\kappa$ & $ a \, (\rm fm)$   \\ \hline
   & $1.27$   & $-1.37$   & $-44.16$     & $-0.40$    & $0.75$ \\
 \hline\hline
\end{tabular}
\end{table}

The proposed nucleus-nucleus potential is based on the frozen
density approximation. The time-dependent Hartree-Forck (TDHF)
calculations show that the nucleus-nucleus potential depends on the
incident energy at energies close to the Coulomb barrier and when
the center-of-mass energy is much higher than the Coulomb barrier
energy, potentials deduced with the microscopic theory identify with
the frozen density approximation \cite{Kou08}. We test  the modified
Woods-Saxon potential for the description of heavy-ion elastic
scattering at energies much higher than the Coulomb barrier, since
the reaction time is relatively short and the frozen density
approximation seems to be applicable at these energies. Based on the
optical model, we solve the Schroedinger equation for a given
nucleus-nucleus potential using the traditional Numerov method to
obtain the partial-wave scattering matrix that is used to describe
the elastic scattering data \cite{Yang05}. The real and imaginary
parts of the optical potential adopted in the calculations are
described by the modified Woods-Saxon potential.

We have calculated the elastic scattering angular distributions for
the reactions $^{12}$C+$^{208}$Pb, $^{16}$O+$^{208}$Pb,
$^{12}$C+$^{90}$Zr and $^{16}$O+$^{63}$Cu at different laboratory
energies. The calculated results (solid curves) are shown in Fig. 1,
and the corresponding experimental data (squares)
\cite{San01,Sam86,Vid77,Ball75,Lich94} are also presented for
comparison. The experimental data of the four reactions at different
energies are reasonably well reproduced by the modified Woods-Saxon
potential in which the potential parameters are fixed.

\begin{figure}
\includegraphics[angle=0,width=1.0\textwidth]{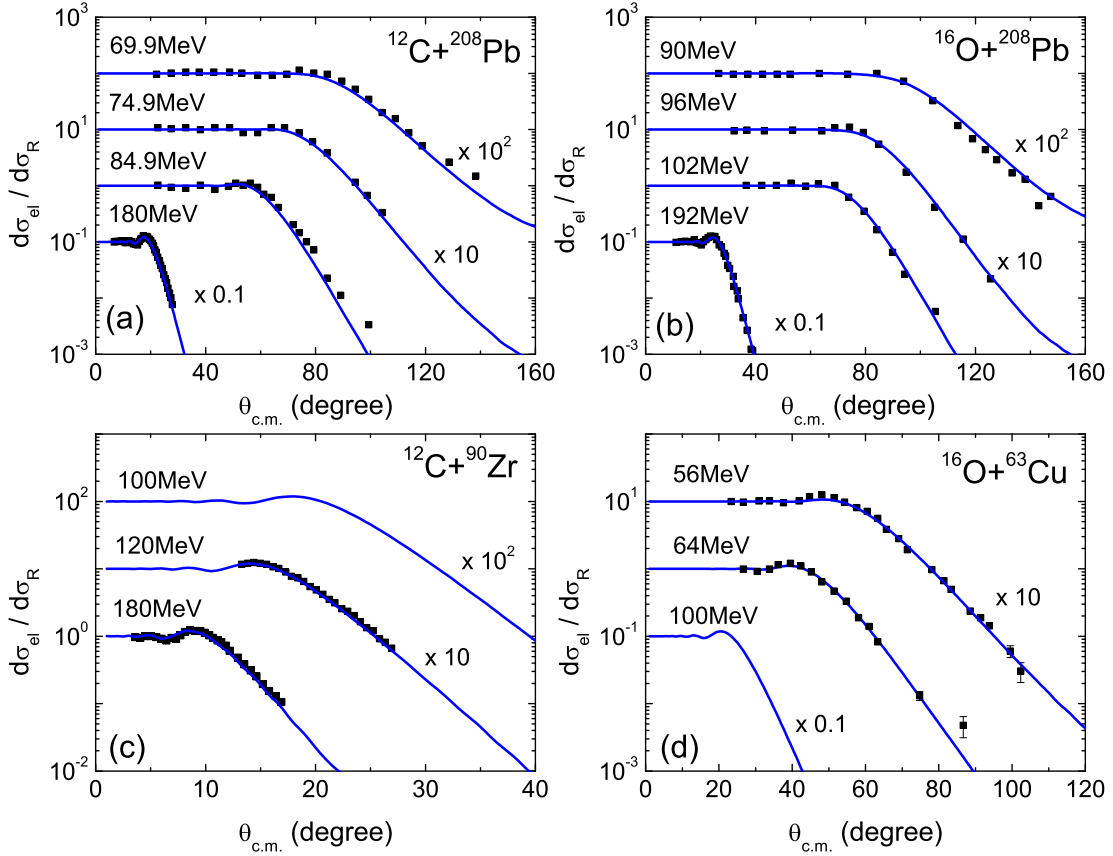}
\caption{(Color online) Elastic scattering angular distributions for
the reactions $^{12}$C+$^{208}$Pb, $^{16}$O+$^{208}$Pb,
$^{12}$C+$^{90}$Zr and $^{16}$O+$^{63}$Cu at different laboratory
energies. The solid curves and the squares denote the calculated
results with the modified Woods-Saxon potential and the experimental
data, respectively. The experimental data are taken from
Refs.\cite{San01,Sam86,Vid77,Ball75,Lich94}.}
\end{figure}

We further test the MWS potential for the description of heavy-ion
fusion at above barrier energies. At these energies, the
 fusion cross section is usually described by the classical formula
\begin{eqnarray}
\sigma_{\rm fus}(E_{\rm c.m.}) =  \pi R_{f}^2 \left ( 1 - B/E_{\rm
c.m.} \right )
\end{eqnarray}
with the fusion radius $R_{f}$ and the height of the fusion barrier
$B$. Fig. 2 shows the fusion excitation function of
$^{16}$O+$^{208}$Pb. Taking $B$ to be the barrier height $B_0$
(78.72 MeV) of the modified Woods-Saxon potential, the fusion cross
sections at above barrier energies can not be reproduced by the
Eq.(5) (see the dash-dotted curve in Fig. 2). In order to describe
the fusion cross sections satisfactorily, we introduced an empirical
barrier distribution to take into account the multi-dimensional
character of a realistic barrier due to the coupling to internal
degrees of freedom of the binary system in our previous paper
\cite{Liu06}. We proposed an effective weight function for
describing the barrier distribution,
\begin{eqnarray}
D_{\rm eff}(B)= \left\{
\begin{array} {r@{\quad:\quad}l}
D_1(B)& B<B_x \\
D_{\rm avr}(B)& B \ge B_x
\end{array} \right.
\end{eqnarray}
where $D_{\rm avr}(B)=(D_1(B)+D_2(B))/2$ and $B_x$ is the left cross
point of $D_1(B)$ and $D_2(B)$. $D_1(B)$ and $D_2(B)$ are two
Gaussian functions \cite{Liu06,Wang08} which depend on the barrier
height $B_0$ of the modified Woods-Saxon potential. The effective
weight function $D_{\rm eff}$ of the reaction $^{16}$O+$^{208}$Pb is
shown in the sub-figure of Fig.2. Taking $B$ to be the most probable
barrier height $B_{\rm m.p.}$ (74.43 MeV) according to the $D_{\rm
eff}$, the fusion cross sections at above barrier energies are
reproduced reasonably well (see the solid curve in Fig.2). With the
empirical barrier distribution, the fusion cross sections and the
mean barrier heights of a large number of reactions can be
reproduced well \cite{Liu06,Wang07,Wang08}.

\begin{figure}
\includegraphics[angle=0,width=0.8\textwidth]{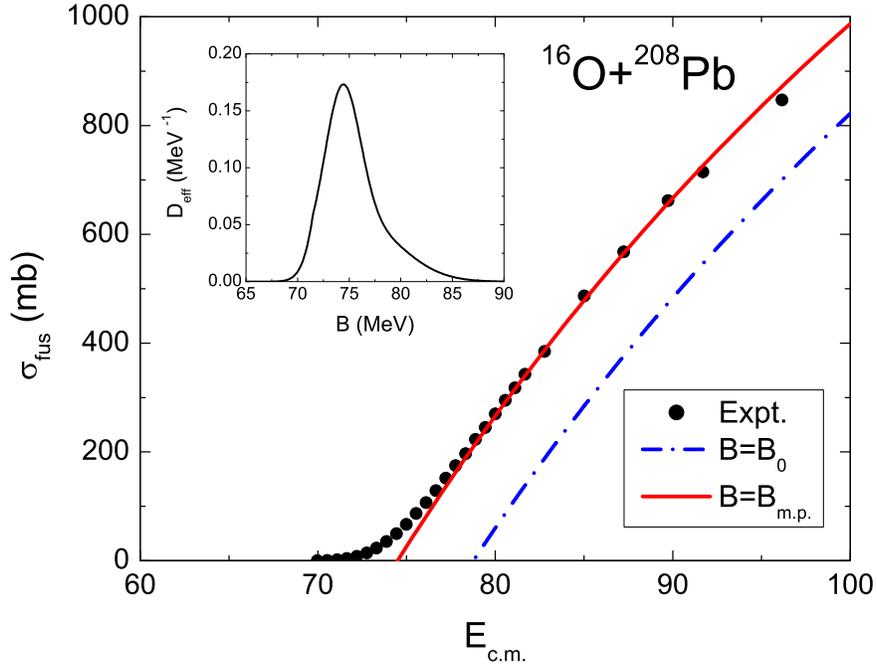}
\caption{(Color online) Fusion excitation function of the reaction
$^{16}$O+$^{208}$Pb. The dash-dotted and solid curves denote the
calculated results with Eq.(5) by taking $B=B_0$ and $B=B_{\rm
m.p.}$, respectively. The solid circles denote the experimental data
\cite{Mor99}. The sub-figure shows the effective weight function of
the reaction.}
\end{figure}

From the above discussion, one finds that for the heavy-ion elastic
scattering at above barrier energies the modified Woods-Saxon
potential which is based on the frozen density approximation gives
nice results. But the fusion cross section of the same reaction
system can not be described well with the potential and the barrier
distribution needs to be introduced to reproduce the fusion data. In
recently published paper \cite{Bas08}, the authors proposed two
optical potentials for describing the reactions $^{12}$C+$^{208}$Pb
and $^{16}$O+$^{208}$Pb, respectively.  Both the elastic scattering
and fusion data can be satisfactorily described with the potentials
at energies around the Coulomb barrier. At energies much higher than
the Coulomb barrier (for example, $E_{\rm lab}=192$ MeV for
$^{16}$O+$^{208}$Pb), the elastic scattering data can not be
reproduced well by the potential. In this work, we aim to find a
nucleus-nucleus potential for describing the reactions
systematically.

\begin{center}
\textbf{B.  Description of Large-angle Quasi-elastic Scattering}
\end{center}

As a good counterpart of the fusion reaction, the large-angle
quasi-elastic scattering is studied to explore the nucleus-nucleus
potential. In this work, we explore the influence of the empirical
barrier distribution proposed for the fusion reactions on the
large-angle quasi-elastic scattering.

It is thought that the quasi-elastic differential cross section can
be expressed as a weighted sum of the eigenchannel elastic
differential cross sections under the adiabatic and iso-centrifugal
approximation \cite{And88,Zhang98}. Similar to the description of
fusion with the empirical barrier distribution, we describe the
large-angle quasi-elastic scattering cross section with the
effective weight function $D_{\rm eff}(B)$ at energies around the
Coulomb barrier,
\begin{eqnarray}
\frac{d\sigma_{\rm qel}}{d\sigma_{R}}(E_{\rm c.m.})=P_{\rm
eff}+P_{\rm corr},
\end{eqnarray}
with
\begin{eqnarray}
P_{\rm eff}=\frac{1}{F_0}\int_{0}^{ \infty }D_{\rm
eff}(B)\frac{d\sigma_{\rm el}}{d\sigma_{R}}(E_{\rm c.m.},B)dB,
\end{eqnarray}
and $P_{\rm corr}$ is a small correction term. $\frac{d\sigma_{\rm
el}}{d\sigma_{R}}$ is the ratio of the elastic cross section
$\sigma_{\rm el}$ to the Rutherford cross section $\sigma_{R}$.
$F_0$ is a normalization constant $F_0=\int D_{\rm eff}(B)dB$.
Within the semi-classical perturbation theory, a semi-classical
formula for the backward scattering ($\theta=\pi$) is given
\cite{Land81,Hag04},
\begin{equation}
\frac{d\sigma_{\rm el}}{d\sigma_R}(E_{\rm c.m.},B) =\left (
1+\frac{V_N(R_c)}{E_{\rm c.m.}}\, \sqrt{\frac{ Z_1 Z_2 e^2 }{E_{\rm
c.m.} }\frac{\pi}{a} } \, \right ) \cdot
\frac{\exp\left[-\frac{2\pi}{\hbar\omega}(E_{\rm c.m.}-B)\right]}
{1+\exp\left[-\frac{2\pi}{\hbar\omega}(E_{\rm c.m.}-B)\right]}.
\end{equation}
Where the nuclear potential $V_N(R_c)$ is evaluated at the Coulomb
turning point,
\begin{equation}
V_N(R_c)=\left (B-\frac{ Z_1 Z_2 e^2 }{R_f }\right ) \left (
\frac{1+\exp[(R_f-R_0)/a]}{1+\exp[(R_c-R_0)/a]} \right ),
\end{equation}
with the distance of the closest approach between two nuclei $R_c=
Z_1 Z_2 e^2 /E_{\rm c.m.}$. $a$ is the diffuseness parameter of the
nuclear potential. $Z_1$, $ Z_2$, $E_{\rm c.m.}$ denote the charge
numbers of the projectile and target nuclei and the center-of-mass
energy, respectively. $R_f$ and $\hbar\omega$ are the barrier
position and curvature of the modified Woods-Saxon potential,
respectively.

The correction term $P_{\rm corr}$ in Eq.(7) takes into account some
effects in the quasi-elastic scattering that are not involved in the
empirical barrier distribution (which was proposed for describing
fusion reactions). In this work, we assume that the correction term
mainly comes from nucleons transfer. In principle, the transfer
process also affects the fusion process and the effect of nucleons
transfer may have been implicitly taken into account in the
empirical barrier distribution. However the influence of nucleons
transfer on the quasi-elastic scattering may differ from the
influence on the fusion process, and a small correction term seems
to be required.

For the quasi-elastic scattering, we first investigate the
dependence of the transfer probabilities on the incident energies.
The transfer probabilities $P_{\rm tr}$ can be written as $P_{\rm
tr}=(d\sigma_{\rm tr}/d\Omega)/(d\sigma_{\rm R}/d\Omega)$, where
$d\sigma_{\rm tr}/d\Omega$ is the transfer cross section
\cite{Rehm91}. At energies below the barrier the transfer
probability increases with increasing incident energies because the
distance of closest approach becomes small, leading to an increase
in the nuclear overlap. At above barrier energies, on the contrary,
the transfer probability decreases with increasing energies since
the increased overlap results in more dissipative collisions which
finally results in fusion \cite{Shr00}. For the transfer at energies
below the barrier, we describe the transfer probability using the
traditional semi-classical method which is mentioned previously in
the introduction section. In this work, only the one-neutron
transfer channels are taken into account for simplicity in the
calculation of the transfer probability at sub-barrier energies. For
the transfer at energies above the barrier, we find the transfer
probabilities are close to the Gaussian function $D_2$ of the
empirical barrier distribution. For example, in Fig. 3 the measured
transfer probabilities for the reaction $^{16}$O+$^{232}$Th are
compared with the corresponding Gaussian function $D_2$ of this
reaction at energies above the barrier. The experimental data are in
good agreement with $D_2$ at above barrier energies.

\begin{figure}
\includegraphics[angle=0,width=0.8\textwidth]{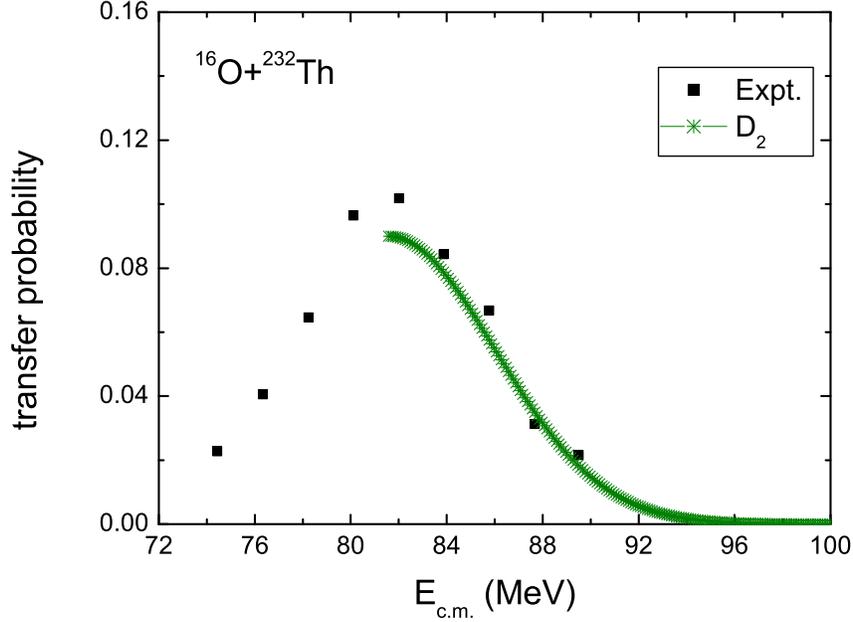}
 \caption{(Color online) Transfer probability of
$^{16}$O+$^{232}$Th. The squares denote the measured transfer
probabilities \cite{Shr00} including the channels of 1p, 1p1n,
1$\alpha$, 2p and 2p1n transfers to the target nuclei.  The crossed
curve denotes the Gaussian function $D_2$ in the empirical barrier
distribution.}
\end{figure}

Based on the above discussion, we write the transfer probability:
\begin{eqnarray}
P_{\rm tr}(E_{\rm c.m.})= f \left\{
\begin{array} {r@{\quad:\quad}l}
P_0\exp(-2 \alpha  R_c)& E_{\rm c.m.}\le B_{\rm m} \\
D_{2}(E_{\rm c.m.}) & E_{\rm c.m.} \ge B_{\rm h} \\
\left ( 1+(F-1)\frac{B_{\rm h}-E_{\rm c.m.}}{B_{\rm h}-B_{\rm m}}
\right ) D_2(E_{\rm c.m.}) & B_{\rm m} <E_{\rm c.m.}<B_{\rm h}

\end{array} \right.
\end{eqnarray}
with the strength factor $f=1$ MeV. $P_0$ is a normalization
constant given by $P_0=D_2(B_0)/\exp(-2 \alpha Z_1 Z_2 e^2/B_0)$
with the transfer form factor $\alpha=\sqrt{ 2 \mu E_b/\hbar^2}$.
$\mu$ is the reduced mass of the transferred nucleons, $E_b$ is the
effective binding energy of the transferred nucleons \cite{Bona99}.
$B_0$ is the barrier height of the modified Woods-Saxon potential,
$B_{\rm m}$ is the mean barrier height of the barrier distribution,
\begin{eqnarray}
B_{\rm m}=\frac{\int B \; D_{\rm eff}(B)\; dB}{\int D_{\rm eff}(B)\;
dB}.
\end{eqnarray}
We take $B_{\rm h}=2 B_0 -B_{\rm m}$ in this work. In order to have
a smooth function for the transfer probability $P_{\rm tr}$ from
sub-barrier energies to above barrier energies, we introduce a
function for $P_{\rm tr}$ in the energy region $B_{\rm m} <E_{\rm
c.m.}<B_{\rm h}$ with a factor $F=\frac{P_0}{D_2(E_{\rm
c.m.})}\exp(-2 \alpha \, Z_1 Z_2 e^2/B_{\rm m})$ to link the two
functions $P_0\exp(-2 \alpha  R_c)$ and $D_2$ describing the $P_{\rm
tr}$ at sub-barrier and at above barrier energies, respectively.

We assume $P_{\rm corr} \approx P_{\rm tr}$. Both the fusion and
quasi-elastic scattering cross sections of a series of reactions
have been studied with the proposed approach in this work. Fig. 4 to
Fig. 6 show the calculated quasi-elastic scattering and fusion cross
sections for the reactions $^{16}$O+$^{144}$Sm, $^{16}$O+$^{154}$Sm,
$^{16}$O+$^{92}$Zr, $^{16}$O+$^{186}$W, $^{32}$S+$^{208}$Pb and
$^{16}$O+$^{116}$Sn. The experimental data
\cite{Tim95,Lei95,New01,Timm98,Pia02,Van01} are also presented for
comparison. The solid circles and squares denote the measured fusion
cross sections $\sigma_{\rm fus}$ and large-angle quasi-elastic
scattering cross sections, respectively. The solid curves in (a) and
(c) of Fig. 4 to Fig. 6 denote the calculated results for
$\sigma_{\rm fus}$ with the proposed empirical barrier distribution
(see details in Refs.\cite{Liu06,Wang08}). The crossed curves in (b)
and (d) denote the calculated quasi-elastic scattering cross
sections with Eq.(7). The dashed curves denote the results for
$P_{\rm eff}$, i.e. the contribution of the empirical barrier
distribution to the quasi-elastic scattering. We find that both the
fusion excitation functions and the quasi-elastic scattering
excitation functions of the six reactions can be satisfactorily well
reproduced. In Fig. 7 we compare the measured quasi-elastic
scattering excitation functions (squares) of the reactions
$^{12}$C+$^{142}$Nd \cite{Maj03}, $^{16}$O+$^{232}$Th \cite{Shr00},
$^{16}$O+$^{64}$Zn \cite{Hui07} and $^{32}$S+$^{110}$Pd \cite{Cap00}
with the calculated results with Eq.(7) (crossed curves). The
calculated quasi-elastic scattering cross sections of the four
reactions $^{12}$C+$^{142}$Nd, $^{16}$O+$^{232}$Th,
$^{16}$O+$^{64}$Zn and $^{32}$S+$^{110}$Pd are in good agreement
with the experimental data. Fig. 4 to Fig. 7 indicate that the
modified Woods-Saxon potential together with the empirical barrier
distribution can simultaneously describe the quasi-elastic
scattering and fusion of a number of reactions reasonably well.

\begin{figure}
\includegraphics[angle=0,width=1.0\textwidth]{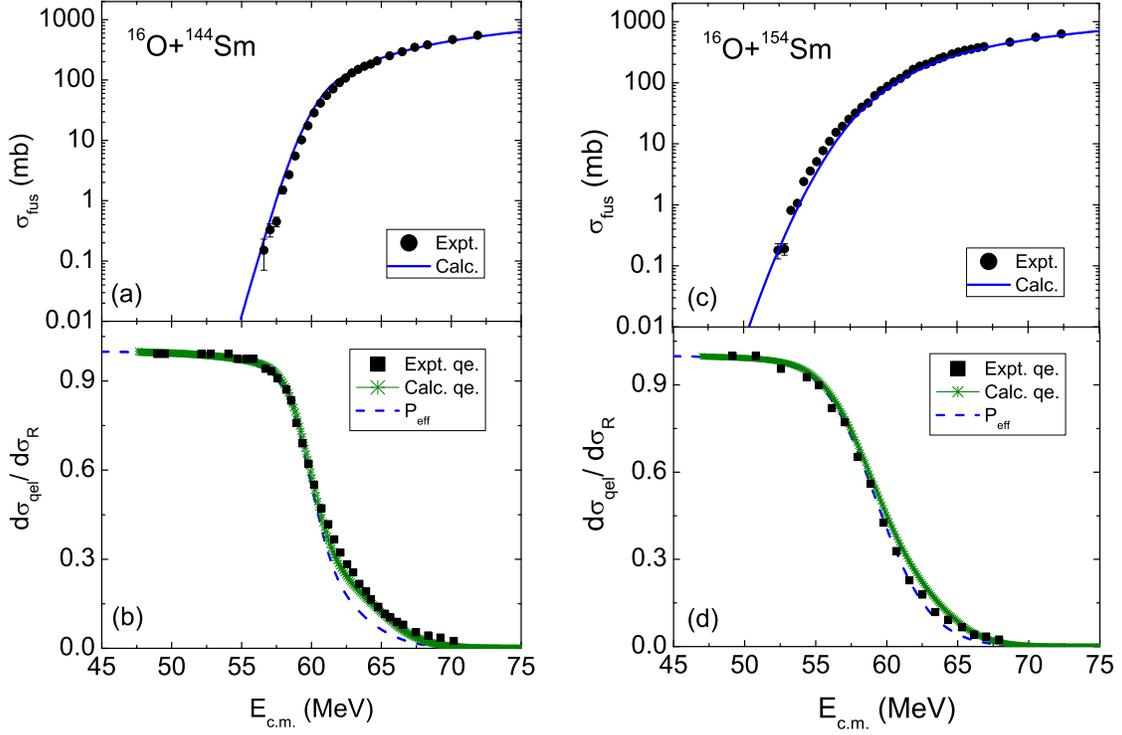}
\caption{(Color online) Fusion cross sections and quasi-elastic
scattering cross sections as a function of energy for the reactions
$^{16}$O+$^{144}$Sm and $^{16}$O+$^{154}$Sm. The solid circles and
squares denote the measure fusion cross sections $\sigma_{\rm fus}$
and quasi-elastic scattering cross sections, respectively. The solid
curves in (a) and (c) denote the calculated results for $\sigma_{\rm
fus}$. The crossed curves in (b) and (d) denote the calculated
results with Eq.(7). The dashed curves denote the results for
$P_{\rm eff}$. The experimental data of fusion and quasi-elastic
scattering are taken from Ref.\cite{Lei95} and \cite{Tim95},
respectively.}
\end{figure}

\begin{figure}
\includegraphics[angle=0,width=1.0\textwidth]{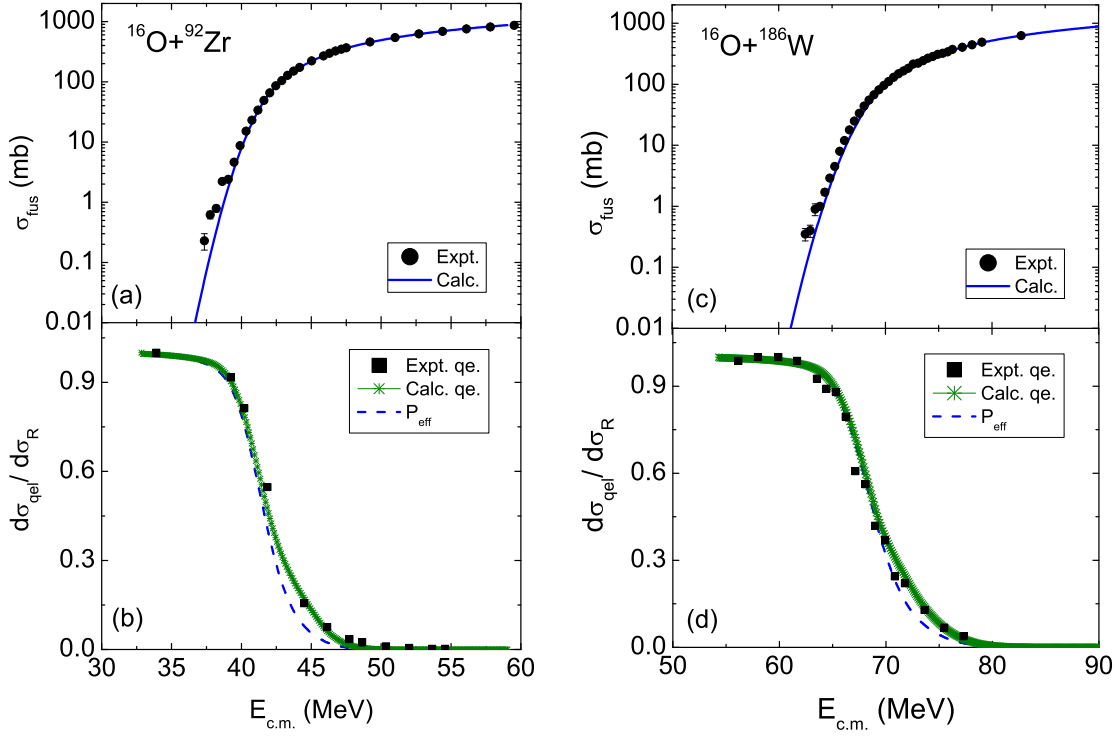}
\caption{(Color online) The same as Fig. 4, but for the reactions
$^{16}$O+$^{92}$Zr and $^{16}$O+$^{186}$W. The experimental data are
taken from Refs.\cite{Tim95,New01,Lei95}.}
\end{figure}

\begin{figure}
\includegraphics[angle=0,width=1.0\textwidth]{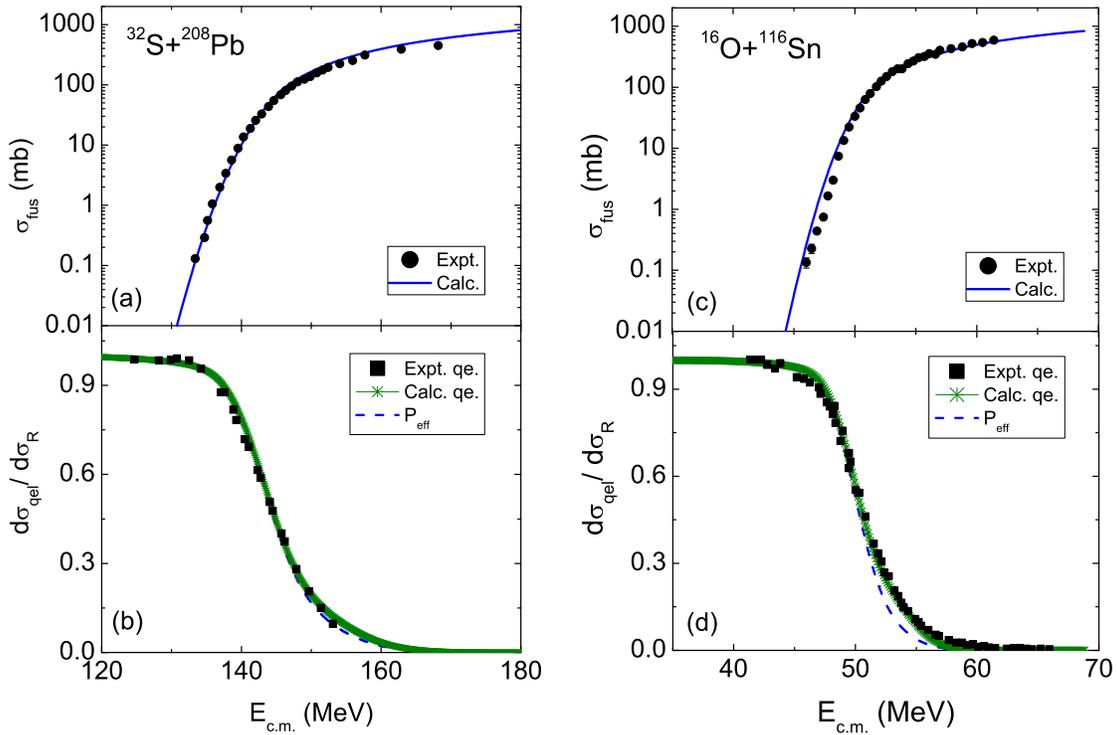}
\caption{(Color online) The same as Fig. 4, but for the reactions
$^{32}$S+$^{208}$Pb and $^{16}$O+$^{116}$Sn. The experimental data
are taken from Refs.\cite{Timm98,Pia02,Van01}.}
\end{figure}

\begin{figure}
\includegraphics[angle=0,width=1.0\textwidth]{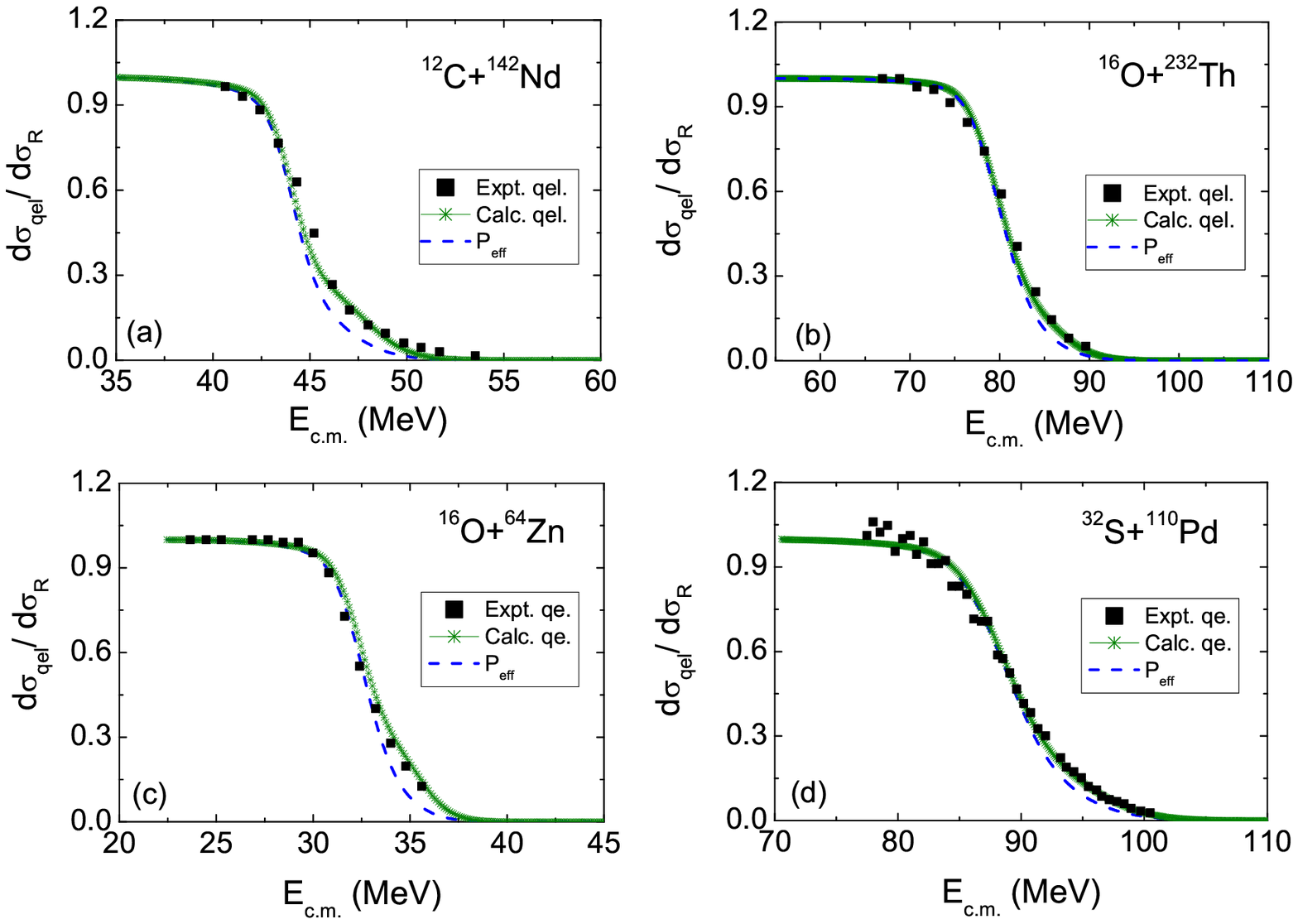}
\caption{(Color online) Quasi-elastic scattering cross sections as a
function of energy for the reactions $^{12}$C+$^{142}$Nd,
$^{16}$O+$^{232}$Th, $^{16}$O+$^{64}$Zn and $^{32}$S+$^{110}$Pd. The
squares and the crossed curves denote the measured and calculated
quasi-elastic scattering cross sections, respectively. The dashed
curves denote the calculated results for $P_{\rm eff}$.}
\end{figure}

\bigskip
\bigskip

\begin{center}
\textbf{III. CONCLUSION AND DISCUSSION}
\end{center}

In this work, we have studied the heavy-ion elastic and large-angle
quasi-elastic scattering with the same nucleus-nucleus potential
proposed for describing fusion reactions. The elastic scattering
angle distributions of a series of reactions at energies much higher
than the Coulomb barrier can be reasonably well reproduced by the
modified Woods-Saxon potential which is based on the frozen density
approximation systematically. With the same potential the fusion
cross sections at above barrier energies can not be reproduced and
an empirical barrier distribution (which takes into account the
coupling of other degrees of freedom) is required to reproduce the
fusion data. It seems that the coupling of other degrees of freedom
to the relative motion of the nuclei is obvious in heavy-ion fusion
processes whereas the frozen density approximation is applicable in
the elastic scattering process at energies much higher than the
Coulomb barrier. With the empirical barrier distribution function
based on the modified Woods-Saxon potential, the fusion cross
sections of a series of reactions have been well reproduced.
Further, with the same empirical barrier distribution and taking
into account the correction term that mainly comes from the nucleons
transfer, the calculated large-angle quasi-elastic scattering cross
sections of these reactions are in good agreement with the
experimental data.

\begin{center}
\textbf{ACKNOWLEDGEMENTS}
\end{center}
One of the authors (N.Wang) is grateful to Prof. Yongxu Yang for
providing a code to calculate elastic scattering angular
distribution and for fruitful discussions. This work is supported by
National Natural Science Foundation of China, No. 10465001.

\end{document}